\documentclass[a4paper,reqno]{amsart}
\usepackage{enumerate,amsmath,amsthm,amsfonts}
\usepackage{amsmath}
\usepackage{amsfonts}
\usepackage{amssymb}

\usepackage{epsfig}
\usepackage{graphicx,color}
\usepackage{graphpap,color}
\usepackage{graphicx}
\usepackage{latexsym}
\usepackage{float}
\usepackage{setspace}
\usepackage[linesnumbered,ruled,vlined]{algorithm2e}
\usepackage{lineno}     
\usepackage{hyperref}

\def\refname {REFERENCES}
\newtheorem{theorem} {Theorem}[section]

\newtheorem{proposition} [theorem]{Proposition}
\def\abstractname {\bf Abstract}

\def\figurename {Figure}

.360pk
.360pk
.360pk

\begin{document}
\include {mak}
~\vspace{-16mm}

\newcommand{\R}{\mathbb{R}}
\newcommand{\Z}{\mathbb{Z}}
\newcommand{\cH}{\mathcal{H}}

\newcommand{\norm}[1]{\left\|#1\right\|}
\newcommand{\hkd}[1]{\left\langle#1\right\rangle}
\newcommand{\krk}[1]{\left\{#1\right\}}
\newcommand{\krb}[1]{\left(#1\right)}
\newcommand{\abs}[1]{\left|#1\right|}
\newcommand{\ol}[1]{\overline{#1}}
\newcommand{\rref}[1]{[\ref{#1}]}

\oddsidemargin 16.5truemm
\evensidemargin 16.5truemm

\thispagestyle{plain}

\vspace{-0.25cc}

\vspace{1.2cc}

\vspace{1.5cc}

\begin{center}
{\Large\bf A MORE PRECISE ELBOW METHOD FOR OPTIMUM K-MEANS CLUSTERING\\ 
\vspace{1.5cc}
{\large\sc Indra Herdiana$^{1}$, M Alfin Kamal$^{2}$, Triyani $^{3}$, Mutia Nur Estri $^{4}$, and Renny $^{5}$}\\

\vspace{0.3 cm} {\small $^{1}$Jenderal Soedirman University, Indonesia, indra.herdiana@unsoed.ac.id\\ $^{2}$alfin.kamal@mhs.unsoed.ac.id\\
$^{3}$triyani@unsoed.ac.id\\
$^{4}$mutia.estri@unsoed.ac.id\\
$^{5}$renny@unsoed.ac.id}

\rule{0mm}{6mm}
}

\vspace{1.5cc}

\parbox{24cc}{{\Small{\bf Abstract.}
K-means clustering is an unsupervised clustering method that requires an initial decision of number of clusters. One method to determine the number of clusters is the elbow method, a heuristic method that relies on visual representation. The method uses the number based on the elbow point, the point closest to $90^\circ$ that indicates the most optimum number of clusters. This research improves the elbow method such that it becomes an objective method. We use the analytical geometric formula to calculate an angle between lines and real analysis principle of derivative to simplify the elbow point determination. We also consider every possibility of the elbow method graph behaviour such that the algorithm is universally applicable. The result is that the elbow point can be measured precisely with a simple algorithm that does not involve complex functions or calculations. This improved method gives an alternative of more reliable cluster determination method that contributes to more optimum k-means clustering.
}}
\end{center}

\vspace{0.25cc}
\parbox{24cc}{\Small {\it Key words and Phrases}: k-means clustering, elbow method, angle between lines
}

\vspace{1.5cc}

\section{INTRODUCTION}

Data are valuable assets in this digital era, inseparable from peoples' lives. Every decision making process involves data. However, data alone cannot produce valuable information without proper management and interpretation. According to \cite{Nolin2019}, profits are made by data management, not data possession, because data do not have an intrinsic value. Statistics is a branch of mathematics that deals with collection, management, analysis, interpretation, and visualisation of data. It is the main key to data management, providing mathematical and scientific method to managing data. For example, \cite{Priyanti2013} and \cite{Usada2021} develop a data management system for demography and primary health facilities respectively that positively impact the society. These two are just a tiny fraction of the essential roles of statistics for human lives.

One of widely used data management method is clustering. Clustering is a method of grouping data to several clusters such that data points in the same cluster have a maximum similarity, while data points in different clusters have a maximum difference. This method is used in many sectors, such as manufacture, transport, sustainable energy, health, and public policy, since it makes data, that are not necessarily meaningful, become meaningful (SSE \cite{Karaca2018, Negi2021, Solikin2022} ). Despite the importance, no clustering methods that fit all \cite{Oyewole2023}. Interestingly, different clustering methods applied to the same data set may result differently. One of notable clustering methods is k-means clustering.

K-means clustering is a clustering method which principle is minimising the distance between every object within a cluster and the centre of the cluster called a centroid; the used distance is the Euclidean distance \cite{Everitt2011}. In other words, objects are grouped into one cluster with the nearest centroid, indicating the maximum similarity centred in the centroid. Some examples of k-means use in data management for further application can be SSEn in \cite{Sembiring2021, Amelia2023}. One factor that makes this method widely used is its simplicity \cite{Li2017}. K-means clustering uses simple calculations, such as the Euclidean distance, that are easily calculable by both manually and computationally. K-means clusering also converges fast \cite{Li2017}. However, although it converges faster than k-medoids clustering, k-means clustering needs an initial determination of number of clusters. This number must not be determined negligently.

There are several methods to determine such a number of clusters, such as the elbow method, gap statistics, the silhouette method, and canopy \cite{Yuan2019}. These methods have unique advantages, disadvantages, and compatibility. However, despite its simplicity, the elbow method has received criticism for its accuracy. The elbow method determines the number of clusters based on a graph that connects some number of clusters and corresponding sum of squared error (SSE) of each cluster. The optimum number of cluster is determined by the point forming an "elbow", the pointiest one, indicating that higher numbers of clusters do not significantly reduce the SSE. The criticism lies on the fact that this method relies on subjective visual interpretation. The elbow point may be chosen inaccurately, thus affecting the clustering result. Here are some examples of the elbow method implementations.

\begin{figure}[H]
	\centering
	\includegraphics[scale=0.4]{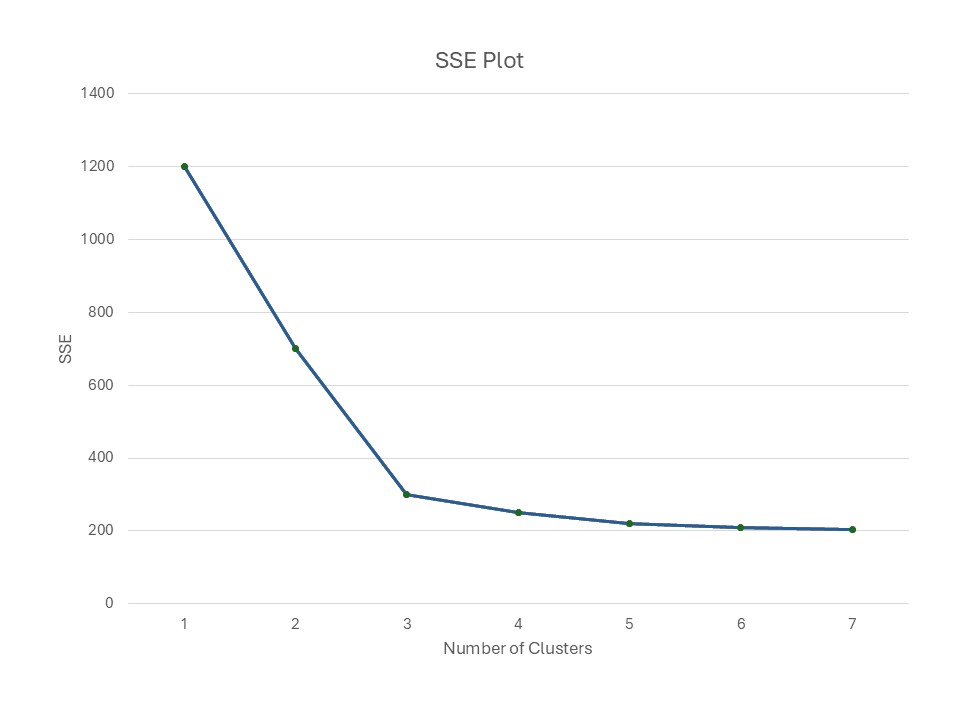}
	\caption{SSE plot with a clear elbow}
	\label{fig:clearelbow}
\end{figure}

\begin{figure}[H]
	\centering
	\includegraphics[scale=0.4]{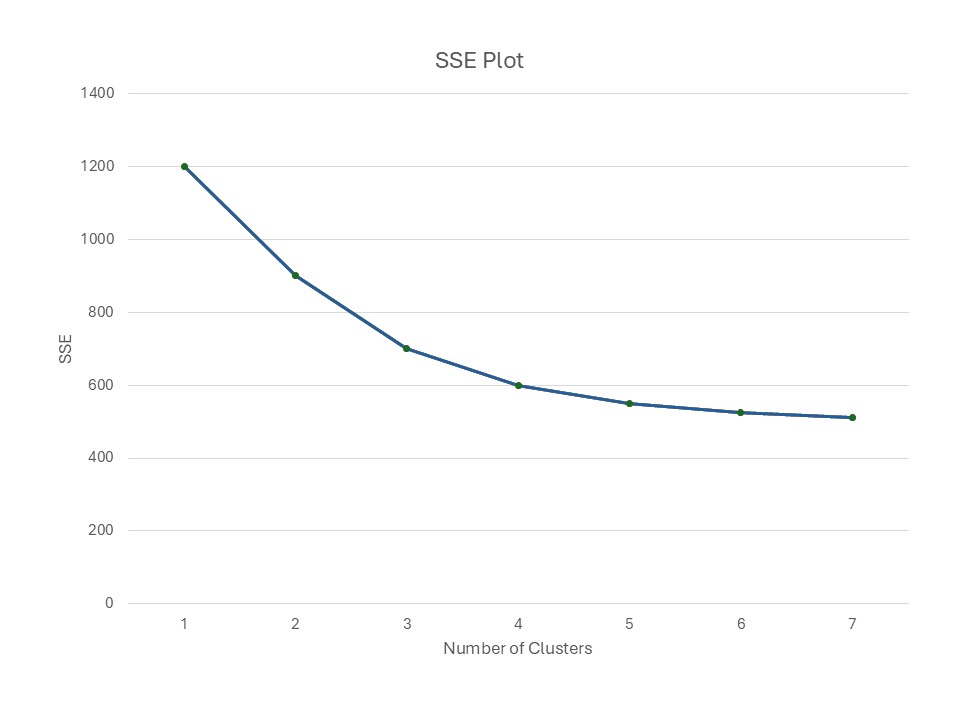}
	\caption{SSE plot with a unclear elbow}
	\label{fig:unclearelbow}
\end{figure}

Figure \ref{fig:clearelbow} illustrates a clear elbow point; we can immediately choose $k=3$ as the optimum number of clusters. However, there may be a case depicted by Figure \ref{fig:unclearelbow} where the elbow point is not clearly shown.

The issue extends beyond this point. Although the elbow points appears to be obvious as shown by \ref{fig:clearelbow}, the "obvious elbow" is possible to be "a false elbow". It is because the scaling of the horizontal and vertical axis are often disproportional. In the case of Figure \ref{fig:clearelbow}, the horizontal axis scale is $0$ to $7$, while the vertical axis scale is $0$ to $1400$. This distortion causes misinterpretations and lead to incorrect choices of optimum number of clusters, thus giving unwanted k-means clustering results.

\begin{figure}[H]
	\centering
	\includegraphics[scale=1]{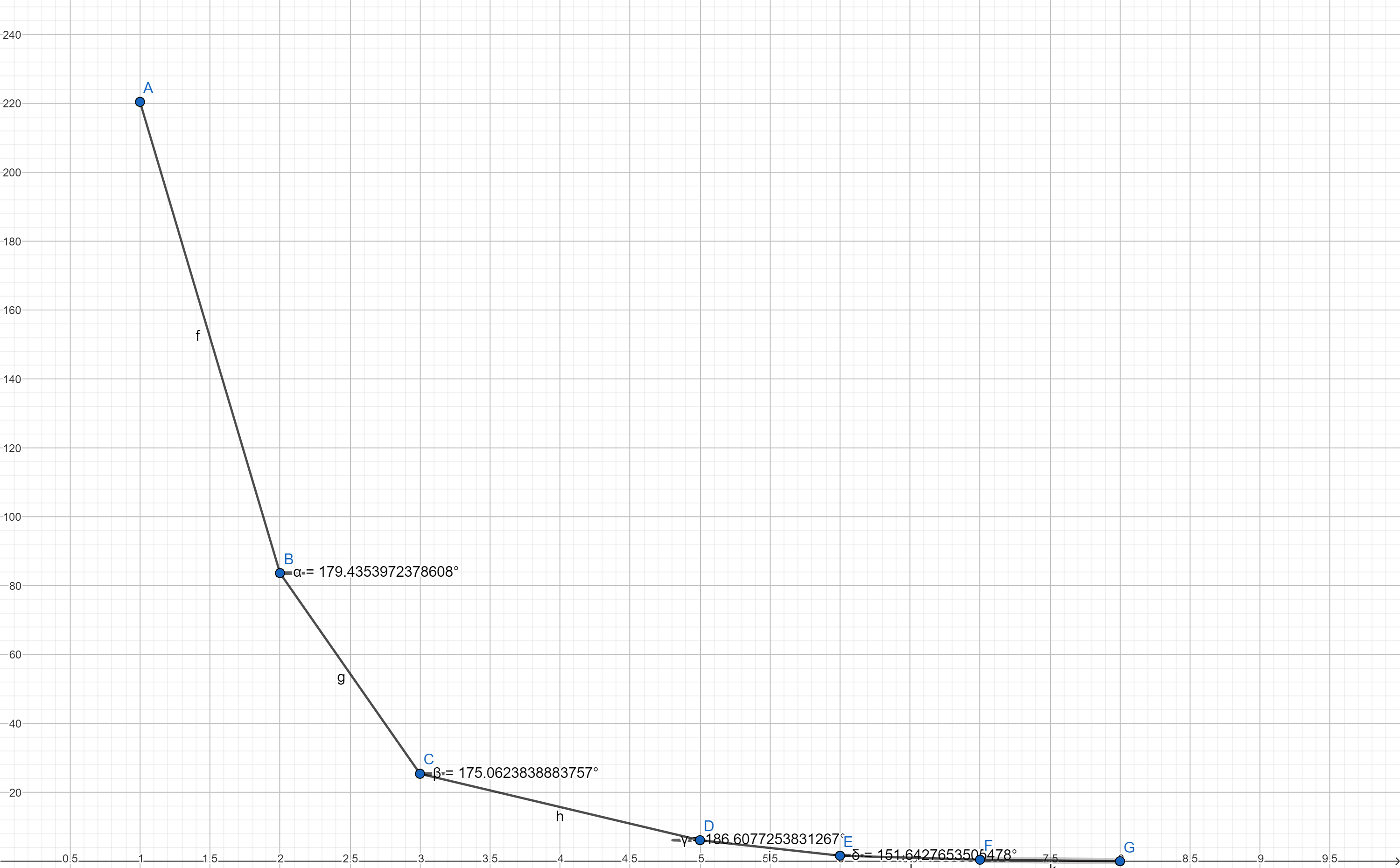}
	\caption{SSE plot with distorted scale}
	\label{fig:elbowfalse}
\end{figure}

\begin{figure}[H]
	\centering
	\includegraphics[scale=0.5]{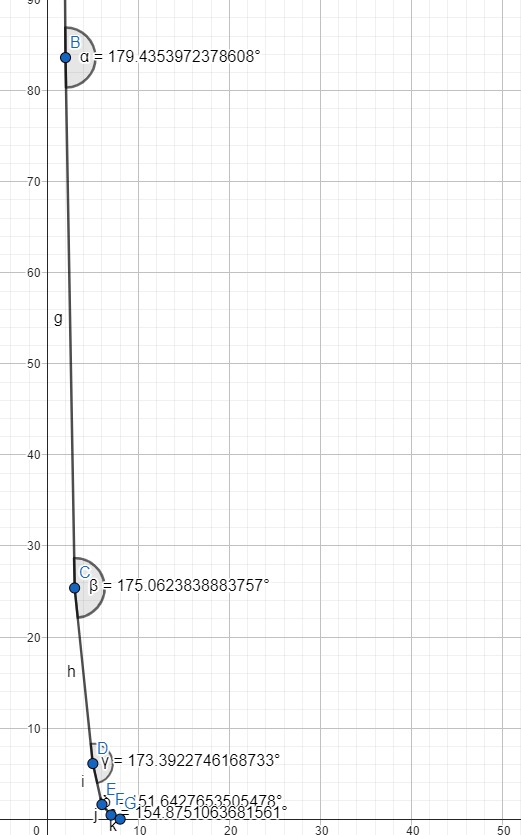}
	\caption{SSE plot with undistorted scale}
	\label{fig:elbowtrue}
\end{figure}

Figure \ref{fig:elbowfalse} shows a distorted plot (that unfortunately often occurs during elbow method processes), whereas Figure \ref{fig:elbowtrue} shows the plot with a proportional scaling. Figure \ref{fig:elbowfalse} gives impression that the optimum number of cluster is $3$, while Figure \ref{fig:elbowtrue} contradicts it.

Due to the complex subjectivity, k-means clustering often involves another method that is considered less subjective, thus neglecting improvisation of the elbow method. Meanwhile, analytical geometry and real analysis have hidden roles in precising this heuristic method. This paper aims to improve the method by changing the subjectivity into objectivity using some principles of analytical geometry and real analysis.

Shi etc. in \cite{Shi2021} provides a method of determining the elbow point precisely using a principle of geometry, namely the cosine rule of a triangle. However, this method involves a long calculation of a triangle side length represented by the distance between two adjacent points. Moreover, this methods uses an inverse trigonometry function that is not a simple standard mathematical operation like addition and multiplication. Sinaga and Yang in \cite{Sinaga2020} create an alternative k-means clustering algorithm where it finds the number of clusters independently without using initial cluster method determination. The method's disadvantage is similar to that of \cite{Shi2021}. It involves a non-standard arithmetic operation namely natural logarithm.

The method proposed in this paper offers an advantage over the methods in \cite{Shi2021, Sinaga2020}. It only involves standard arithmetic operations namely addition, subtraction, multiplication, and division, resulting in more efficient algorithm. Not all programming languages have built-in complex mathematical functions like trigonometric inverses and natural logarithm; even C++ needs to add the package cmath in order to use these operations  \cite{Joe2017}. Therefore, using standard arithmetic functions that are included in all programming languages offer flexibility and simplicity.

The method uses the principle of measuring an angle between two lines formulated in analytical geometry. The real analysis principle of derivative is also used to optimise the algorithm such that non-standard operations occurring in the algorithm can be omitted. Furthermore, the proposed method also considers every possibility of the elbow method graph behaviour such that the elbow point is chosen more precisely. This method can be an alternative to optimise k-means clustering; results in \cite{Shi2021} are even shown to be better than the silhouette method.

\vspace{1.5cc}

\section{MAIN RESULTS}

This section discusses the detailed formulation of determining the elbow point using an analytical geometric approach. Data simulation is also provided using Python.

\subsection{Exact Elbow Point Formulation}

\vspace{.5cc}

The following theories about the elbow method, k-means clustering, and SSE are cited from \cite{Everitt2011, Yuan2019}.

Let $\mathcal{X}$ be a collection of $n$ continuous $p$-dimensional data such. Let $X_i$ denote the $i$-th data point on $\mathcal{X}$ where
\begin{equation*}
	X_i=(x_1,x_2,…,x_p )
\end{equation*}
and $i=1,2,…,n$. Assume that $\mathcal{X}$ does not contain an outlier such that it is suitable for 
k-means clustering. Suppose that we cluster the data set into $k$ clusters, and the determination of $k$ utilises the elbow method.

We initiate the elbow method by plotting the sequence $(SSE(k))$ for $k=1,2,…,n$\footnote{The largest possible number of clusters is the number of data points since it is irrational to have more clusters than data points.} where $SSE(k)$ is the sum of squared errors obtained if the data set is grouped into $k$ clusters by k-means clustering. It is defined by
\begin{equation}
	SSE(k) = \sum_{j=1}^{k} \sum_{X_i \in S_j} \| X_i - C_{S_j} \|^2 \nonumber
\end{equation} 
where $S_j$ is the $j$-th cluster, $C_{S_j}$ is the centroid of cluster $S_j$, and $‖X_i-C_(S_j )‖$ is the Euclidean distance between the data point $X_i$ and its centroid $C_{S_j}$. The SSE plotted here is obtained after iterating the k-means clustering such that no data points change their cluster membership. In other words, the plotted SSE is the most optimum SSE indicated by complete k-means clustering, not the initial SSE calculated when centroids are first-chosen randomly.

K-means clustering groups data points based on the nearest centroid using the Euclidean distance, and centroids change repeatedly by the optimising formula
\begin{equation}
	C_{S_j}=\frac{1}{N(S_j )} \sum_{X_i \in S_j}X_i, \nonumber
\end{equation}
where $N(S_j)$ denotes the number of data points in the cluster $S_j$. The change stops when no data points move to another cluster, thus implying the lowest possible SSE. If the number of clusters increases, the number of centroids also increases. Hence, distances between data points and their nearest centroids tend to shrink, resulting in a smaller SSE. By this deduction, we conclude that the sequence $(SSE(k))$ is monotonically decreasing as greater clusters imply smaller SSEs.

\begin{figure}[H]
	\centering
	\includegraphics[scale=0.4]{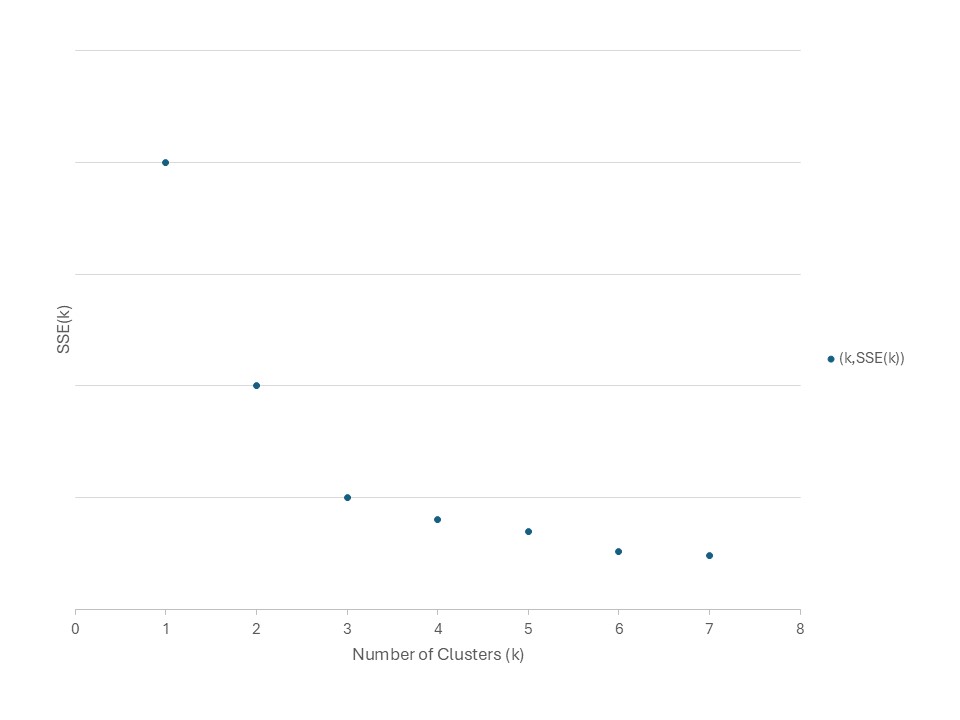}
	\caption{Initial SSE plotting}
	\label{fig:plotSSE}
\end{figure}

The sequence $SSE(k)$, as illustrated by Figure \ref{fig:plotSSE}, is plotted monotonically decreasing. We connect adjacent points by a straight line such that all point (terms of $SSE(k)$) form a continuous function consisting of several straight lines. Every corner of the function graph forms an angle that is used to determine the elbow point.

\begin{figure}[H]
	\centering
	\includegraphics[scale=0.4]{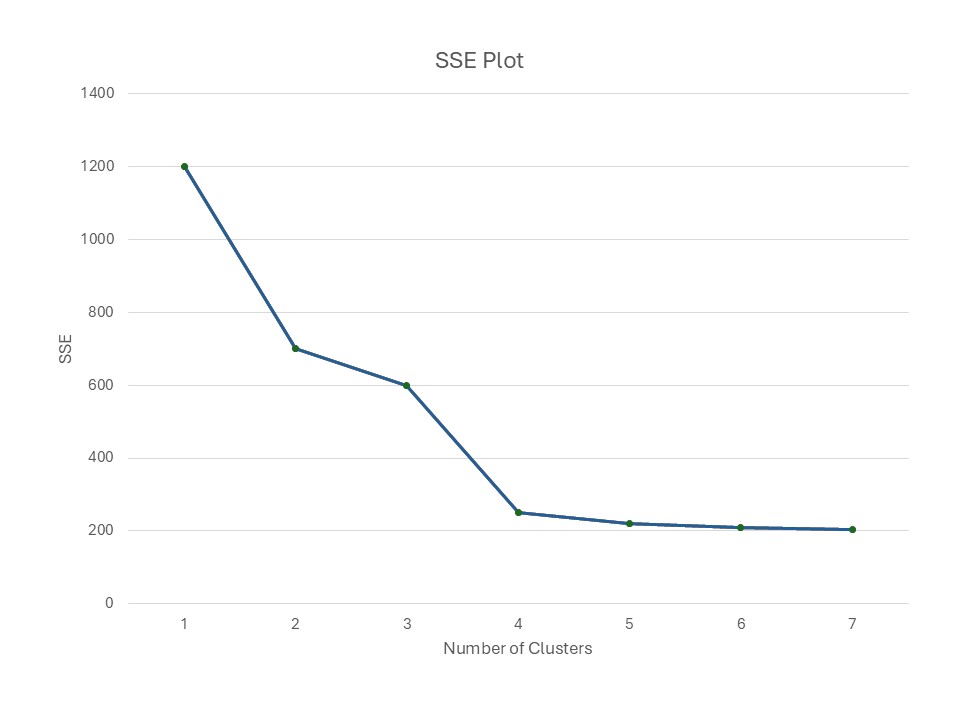}
	\caption{SSE plot with a facing-downwards corner}
	\label{fig:casejeglek}
\end{figure}

The elbow point is the point in which there is no significant drop of SSE afterwards. The typical elbow method determines the angle closest to $90^\circ$ since it indicates the last point with a significant drop of SSE i.e. the graph starts flattening beyond the point. However, since the elbow method, at the beginning, is a heuristic method, the closest-to-$90^\circ$ choice is rather ambiguous. Two disputes need to be addressed. First, the angle measurement is not mathematical. Second, the method appears to ignore the possibility of facing-downwards corners chosen as the elbow point (illustrated by Figure \ref{fig:casejeglek}.

To address the first dispute, we construct a formula to measure the exact angle of the corners. We use the formula of angle between lines that uses line slopes.

\begin{figure}[H]
	\centering
	\includegraphics[scale=0.4]{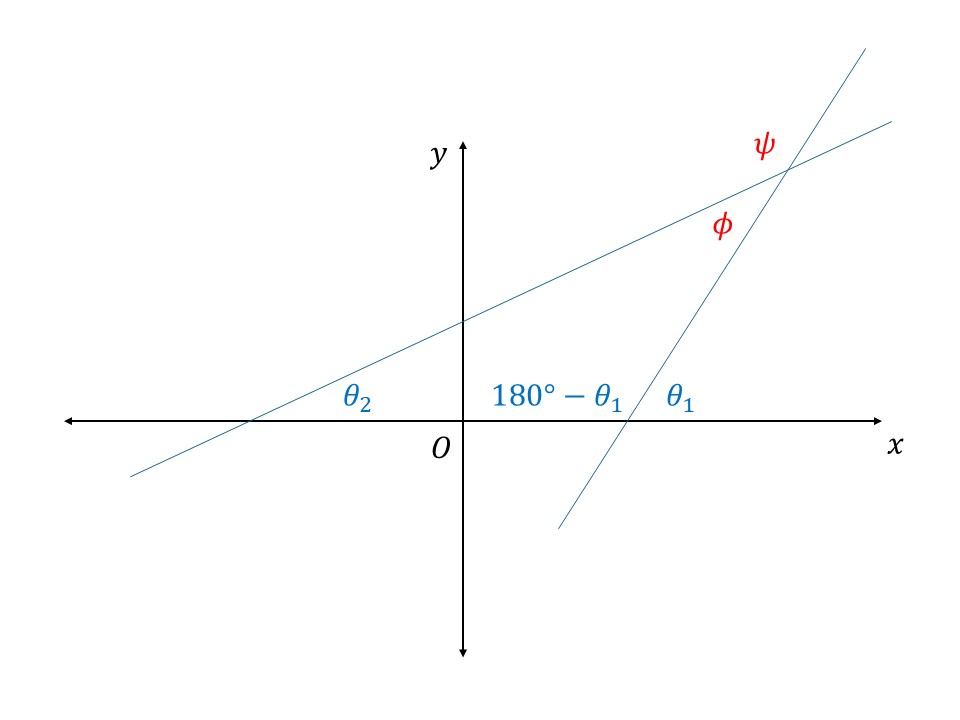}
	\caption{Formula of angles between two lines}
	\label{fig:anglebetweenlines}
\end{figure}

\begin{theorem}
	\label{thm:anglebetweenlines}
	Let $l_1$ and $l_2$ be lines with inclination $\theta_1$ and $\theta_2$ respectively. Let $m_{l_1}$ and $m_{l_2}$ denote the slope of $l_1$ and $l_2$ respectively where $m_{l_1}=\tan{\theta_1}$ and $m_{l_2}=\tan{\theta_2}$. The two lines intersect in a point, forming two supplementary angles, $\phi$ and $\psi$ as illustrated by \ref{fig:anglebetweenlines}. The following holds:
	\begin{equation*}
		\tan{\phi}=\frac{m_{l_1}-m_{l_2}}{1+m_{l_1}m_{l_2}}
	\end{equation*}
	and
	\begin{equation*}
		\tan{\psi}=-\tan{\phi}.
	\end{equation*}
\end{theorem}

The proof is constructed in \cite{Fuller1992}.

Since $\phi$ and $\psi$ are supplementary, unless both are $90^\circ$, the former is located in the first quadrant, and the latter is located in the second quadrant, or vice versa. The angle located in the first quadrant has a positive tangent, whereas the one located in the second quadrant has a negative tangent.

We use the modifed Theorem \ref{thm:anglebetweenlines} to construct angle measurement formula of the SSE graph since the line behaviours are different. However, in this case, we first assume that the second dispute does not exist i.e. the lines forming the graph get more flattened after each term. The flattening condition is indicated by flattening lines i.e. slopes of the lines get closer to $0$ as $k$ increases (the slope of horizontal lines). In other words, we assume that
\begin{equation}
	m_{l_k} > m_{l_{k-1}} \nonumber
\end{equation}
for $k=1,2,...,n$.

\begin{figure}[H]
	\centering
	\includegraphics[scale=0.4]{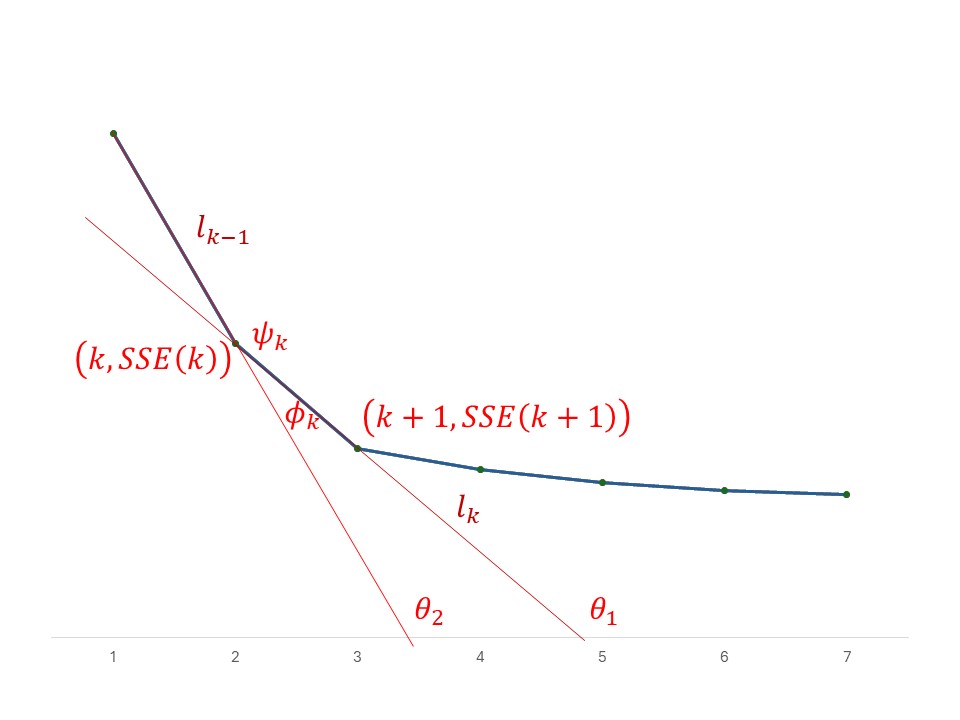}
	\caption{Implementation of Theorem \ref{thm:anglebetweenlines} on SSE plotting}
	\label{fig:maintheorem}
\end{figure}

\begin{theorem}
	\label{thm:elbowangle1}
	 Let $l_k$ denote the straight line connecting the point $(k,SSE(k))$ and $(k+1,SSE(k+1))$, and $\psi_k$ denote the angle of the corner facing upwards. The elbow point is $(k,SSE(k))$ such that it satisfies
	\begin{equation}
	\label{eq:minimumtan}
		\min{(\tan{\psi_k} | k=2,3,...,n-1)}
	\end{equation}
	where
	\begin{equation}
		\tan(\psi_k) = \frac{-SSE(k+1)+2SSE(k)-SSE(k-1)}{1+(SSE(k)-SSE(k-1))(SSE(k+1)-SSE(k))}. \nonumber
	\end{equation}
\end{theorem}

To prove Theorem \ref{thm:elbowangle1}, we first consider the following proposition.

\begin{proposition}
	\label{pro:psi}
	If $\psi$ is the angle that faces upward, we have $90^\circ < \psi < 180^\circ$.
\end{proposition}

\begin{proof}
	If $l_k$ denotes the straight line connecting the point $(k,SSE(k))$ and $(k+1,SSE(k+1))$, since $SSE(k+1)<SSE(k)$ for $k=1,2,...,n$, we have
	\begin{align*}
		m_{l_{k}} &= \frac{SSE(k+1)-SSE(k)}{(k+1)-(k)}\\
		&= SSE(k+1)-SSE(k)\\
		&< 0. 
	\end{align*}
	Moreover, we assume $m_{l_k}>m_{l_{k-1}}$ which implies $m_{l_{k}} - m_{l_{k-1}}>0$. Consequently, since $m_{l_k} m_{l_{k-1}}>0$, we have
	\begin{align*}
		\tan{\phi_k} &= \frac{m_{l_k}-m_{l_{k-1}}}{1+m_{l_k}m_{l_{k-1}}}\\
		&> 0
	\end{align*}
	and
	\begin{align*}
		\tan{\psi}&=-\tan{\phi}\\
		&< 0.
	\end{align*}
	Since $\tan{\psi}<0$, the angle $\psi$ must be in quadrant II or III i.e. $90^\circ < \psi < 180^\circ$ or $180^\circ < \psi < 270^\circ$. However, since $\psi$ and $\phi$ are supplementary, it must not exceed $180^\circ$. Therefore, we have $90^\circ < \psi < 180^\circ$.
\end{proof}

Here is the proof of Theorem \ref{thm:elbowangle1}.

\begin{proof}
	Recall that $l_k$ denote the straight line connecting the point $(k,SSE(k))$ and $(k+1,SSE(k+1))$. As illustrated by Figure \ref{fig:maintheorem}, we have $\theta_1$ and $\theta_2$ are the inclination of $l_{k-1}$ and $l_{k}$ respectively. Since $\psi + (180^\circ - \theta_1) + \theta_2=180^\circ$, we have
	\begin{align*}
		\tan{\phi} &= \tan{(180^\circ - ((180^\circ - \theta_1) + \theta_2) )}\\
		&= \tan{(\theta_1 - \theta_2)}\\
		&= \frac{\tan{\theta_1}-\tan{\theta_2}}{1+\tan{\theta_1}\tan{\theta_2}}\\
		&= \frac{m_{l_k}-m_{l_{k-1}}}{1+m_{l_k}m_{l_{k-1}}}.
	\end{align*}
	
	Elbow point candidates are represented by the angle supplementary to $\phi$, that is $\psi$. Since the angles are supplementary, the angles have opposites tangents i.e.
	\begin{align}
	\label{eq:tanphik}
		&\tan{\psi} = -\tan{\phi} \nonumber\\
	&\iff \tan{\psi} =-\frac{m_{l_k}-m_{l_{k-1}}}{1+m_{l_k}m_{l_{k-1}}}\nonumber\\
	&\iff \tan{\psi} =\frac{m_{l_{k-1}}-m_{l_k}}{1+m_{l_k}m_{l_{k-1}}}.
	\end{align}
	
	The endpoints, $(1,SSE(1)$ and $(n,SSE(n))$, do not form an angle. Therefore, Equation \ref{eq:tanphik} only holds for $k=2,3,...,n-1$.
	
	By the formula of slope of a straight line, since $l_k$ connects $(k,SSE(k))$ and $(k+1,SSE(k+1))$, we have
	\begin{equation*}
		\label{eq:slopelk}
		m_{l_k}=\frac{SSE(k+1)-SSE(k)}{(k+1)-k}
	\end{equation*}
	and
	\begin{equation*}
		\label{eq:slopelk-1}
		m_{l_{k-1}}=\frac{SSE(k)-SSE(k-1)}{k-(k-1)}.
	\end{equation*}
	Therefore, Equation \ref{eq:tanphik} can be written as
	\begin{align*}
			\tan{\psi_k} &= \frac{m_{l_{k-1}}-m_{l_k}}{1+m_{l_k}m_{l_{k-1}}}\\
			&= \frac{\frac{SSE(k)-SSE(k-1)}{(k+1)-k}-\frac{SSE(k+1)-SSE(k)}{k-(k-1)}}{1+(\frac{SSE(k+1)-SSE(k)}{(k+1)-k})(\frac{SSE(k)-SSE(k-1)}{k-(k-1)})}\\
			&= \frac{SSE(k)-SSE(k-1)-(SSE(k+1)-SSE(k))}{1+(SSE(k+1)-SSE(k))(SSE(k)-SSE(k-1))}\\
			&= \frac{-SSE(k-1)+2SSE(k)-SSE(k+1)}{1+(SSE(k+1)-SSE(k))(SSE(k)-SSE(k-1))}.
	\end{align*}
	By Proposition \ref{pro:psi}, we have $90^\circ < \psi < 180^\circ$. Furthermore, since $\tan{\psi}$ is monotonically increasing within the interval\footnote{The derivative of $\tan{\psi}$ with respect to $\psi$ is $\sec^2{\psi}$ that is always positive for $90^\circ < \psi <180^\circ$. Therefore, the function increases.}, the angle $\psi$ closest to $90^\circ$ must be satisfied by the smallest (most negative) $\tan{\psi}$. Therefore, we conclude that the elbow point is $(k,SSE(k))$ that satisfies
	\begin{equation*}
		\min{(\tan{\psi_k} | k=2,3,...,n-1)}.
	\end{equation*}
\end{proof}

\subsection{Alternative Method}
The Theorem \ref{thm:elbowangle1} holds if the second dispute is assumed to be untrue. However, some data sets cause the elbow method graph violating the assumption. In other words, there can be a case that $m_{l_k} \geq m_{l_{k-1}}$.

\begin{figure}[H]
	\centering
	\includegraphics[scale=0.35]{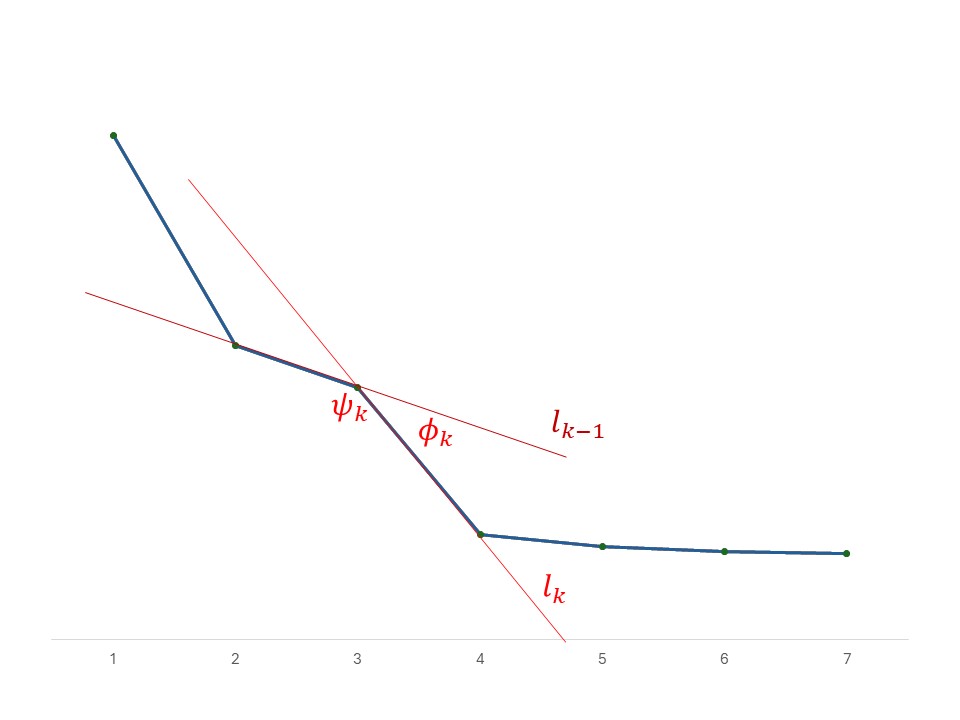}
	\caption{Facing-downwards corner impacts on elbow point determination}
	\label{fig:anglejeglek}
\end{figure}

In Figure \ref{fig:anglejeglek}, the corner that is an intersection of $l_1$ and $l_2$ may be chosen if it is the closest angle to $90^\circ$ if the usual elbow point criteria is used. However, it is clear that the corner must no be the elbow point as there a larger drop of SSE onwards. It does not necessarily indicate the sign of flattening SSE drops.

\begin{proposition}
	\label{pro:ssedrop}
	If $m_{l_k} \leq m_{l_{k-1}}$, then $SSE(k)-SSE(k+1) \geq SSE(k-1)-SSE(k)$.
\end{proposition}

\begin{proof}
	The drop of $SSE(k)$ to $SSE(k+1)$ is $|SSE(k)-SSE(k+1)|$, or for simplicity, since $SSE(k)$ is monotonically decreasing, we have $SSE(k)-SSE(k+1)$ for $k=1,2,...,n$.
	
	Suppose that $m_{l_k} \leq m_{l_{k-1}}$. We then have
	\begin{align}
		\label{eq:ssedrop}
		&m_{l_k} \leq m_{l_{k-1}} \nonumber \\
		&\iff \frac{SSE(k+1)-SSE(k)}{(k+1)-k} \leq \frac{SSE(k)-SSE(k-1)}{k-(k-1)} \nonumber \\
		&\iff SSE(k+1)-SSE(k) \leq SSE(k)-SSE(k-1) \nonumber \\
		&\iff -(SSE(k+1)-SSE(k)) \geq -(SSE(k)-SSE(k-1)) \nonumber\\
		&\iff SSE(k)-SSE(k+1) \geq -SSE(k-1)-SSE(k)
	\end{align}
	The Equation \ref{eq:ssedrop} indicates a higher drop after the point $(k,SSE(k))$.
\end{proof}

\begin{figure}[H]
	\centering
	\includegraphics[scale=0.4]{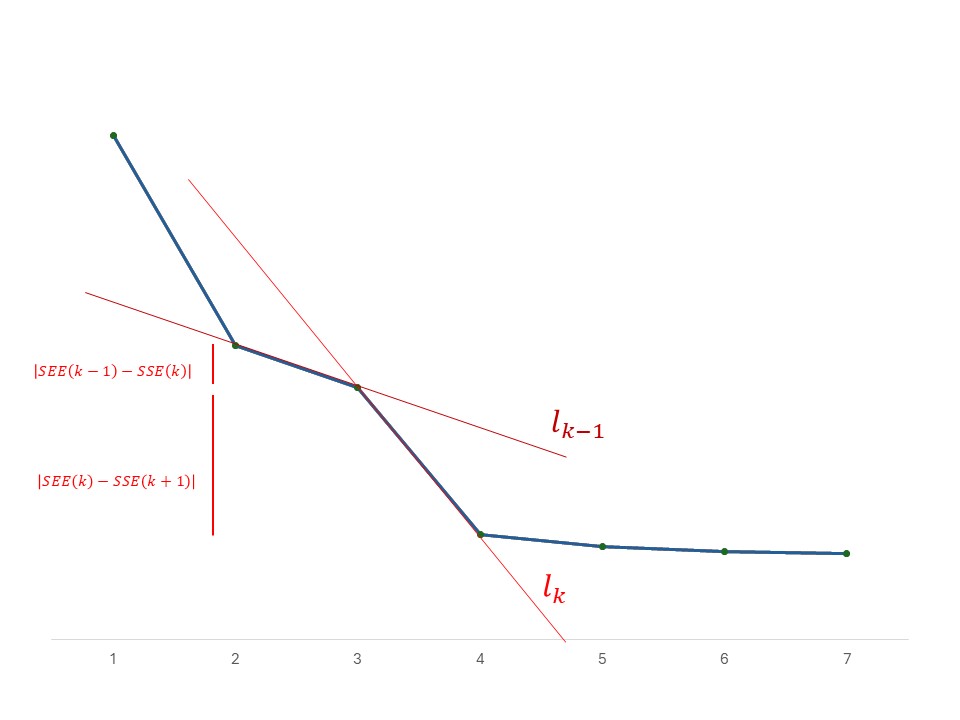}
	\caption{Further Effects of Facing-downwards corners}
	\label{fig:anglejeglekgeneral}
\end{figure}

Figure \ref{fig:anglejeglekgeneral} illustrates the Proposition \ref{pro:ssedrop}.

By discovering the effect of ignoring the second dispute to the elbow point choice, we reconstruct Theorem \ref{thm:elbowangle1} such that it is more universal. We use the same criteria (the closest-to-$90^\circ$) with an additional condition: neglecting the point in Proposition \ref{pro:ssedrop}.

\begin{theorem}
	\label{thm:elbowangle2}
	Suppose that $SSE(k+1) \leq SSE(k)$ for $k=1,2,...,n$. Let $l_k$ denote the straight line connecting the point $(k,SSE(k))$ and $(k+1,SSE(k+1))$, and $\psi_k$ denote the angle of the corner facing upwards. The elbow point is $(k,SSE(k))$ such that it satisfies
	\begin{equation}
		\min{(\tan{\psi_k} | k=2,3,...,n-1; m_{l_k} > m_{l_{k-1}})} \nonumber
	\end{equation}
	where
	\begin{equation}
	\label{eq:tanpsimain}
		\tan(\psi_k) = \frac{-SSE(k+1)+2SSE(k)-SSE(k-1)}{1+(SSE(k)-SSE(k-1))(SSE(k+1)-SSE(k))}.
	\end{equation}
\end{theorem}

Theorem \ref{thm:elbowangle2} adds a condition that is the negation of condition in Proposition \ref{pro:ssedrop} into Equation \ref{eq:minimumtan}. Hence, points indicating a rise of SSE drop, satisfying the condition in Proposition \ref{pro:ssedrop} is neglected.

\subsection{Program Implementation and Simulation}

K-means clustering is often implemented by software, computationally, especially if it involves a large data set. Hence, we also provide an implementation of the constructed formula into computer programming. Here is the pseudo-code.\\

\begin{algorithm}[H]
	\caption{Elbow Method Pseudocode}
	Input a data set consisting of $n$ $m$-dimensional data points $X_i=(x_1,x_2,...,x_p)$\\
	Input array $tanpsi$ with size $n$\\
	Define function $d(X_i,Y_i)$\\
	\quad for i = 1 to p do\\
	\quad \quad sum = sum + $(x_i-y_i)\texttt{\^{}}2$\\
	\quad end for\\
	\quad return sqrt(sum)\\
	Define function $SSE(k)$\\
	\quad do k-means to data set with $k$ cluster(s)\\
	\quad define centroid $C_i$\\
	\quad for i = 1 to k do\\
	\quad \quad sum = sum + $d(X_i,C_i)$\\
	\quad end for\\ 
	\quad return sum\\	
	for k = 2 to n-1 do\\
	\quad if $SSE(k) - SSE(k+1) \geq SSE(k-1) - SSE(k)$ then\\
	\quad \quad $tanpsi[k]$ = 0\\
	\quad else\\
	\quad \quad $tanpsi[k]$ = $(-SSE(k+1)+2SSE(k)-SSE(k-1))$/$(1+(SSE(k)-SSE(k-1))(SSE(k+1)-SSE(k)))$\\
	\quad end if\\
	end for\\
	angle = minimum($tanpsi$)\\
	for k = 2 to n-1 do\\
	\quad if $tanpsi[k]$ = angle then\\
	\quad \quad return k\\
	\quad \quad do k-means clustering with $k$ cluster(s)\\
	\quad \quad break loop\\
	\quad end if\\
	end for
\end{algorithm}

The algorithm begins with defining main variables used for k-means clustering and the elbow method namely a data set and an array containing list of $\tan{\psi}$. We define some functions to simplify the algorithm that are the Euclidean distance and the SSE. The Euclidean distance function involves a looping as a representative of consecutive summation of squared differences; the SSE function also involves a looping of Euclidean distances between every variable and its centroid.

The next step is creating a looping to put $\tan{\psi_k}$ consecutively into the defined array. The formula is of Equation \ref{eq:tanpsimain}. We also put a condition of Theorem \ref{thm:elbowangle2} to ignore corners facing upwards. We put $\tan{\psi_k}=0$ for this kind of points. Hence, the points will not be chosen as the elbow point they will not become the minimum among negative tangents.

Lastly, we define the variable angle as the minimum of $\tan{\psi_k}$. Afterwards, we do a looping to return the number of clusters that results in the minimum $\tan{\psi_k}$. The k-means clustering is then run with the number of clusters.

Some programming languages already include standard functions used in this algorithm, such as the Euclidean distance, SSE, and k-means itself. Python is one oh such programming languages. Here is the implementation in Python for some sample data set (we can change it to different data set).

\begin{algorithm}[H]
	\caption{The Elbow Method Implementation on Python}
	import math\\
	import matplotlib.pyplot as plt\\
	from sklearn.cluster import KMeans\\
	import numpy as np\\
	
	\# Sample 2D data\\
	X = np.array([[1, 1], [1.5, 1.8], [5, 8], [8, 8], [10, 0.6], [9, 11],[0,1],[3,4]])\\
	
	\# Get the number of data points of X\\
	dimension = X.shape[0]\\
	
	\# Calculate WCSS (SSE) for different k values\\
	wcss = []\\
	for i in range(1, dimension+1):  \# Test different values of k (1 to 6 in this example)\\
	kmeans = KMeans(n\_clusters=i, init='k-means++',\\ max\_iter=300, n\_init=10, random\_state=0)\\
	kmeans.fit(X)\\
	wcss.append(kmeans.inertia\_)\\
	
	\# Plot the Elbow Method graph (warning: biased)\\
	plt.plot(range(1, dimension+1), wcss)\\
	plt.title('Elbow Method')\\
	plt.xlabel('Number of clusters')\\
	plt.ylabel('SSE')  \# Within-Cluster Sum of Squares\\
	plt.show()\\
	
	\# Calculate tangent of the angle (tan(psi))\\
	tanpsi = []\\
	tanpsi.append(0)  \# Boundary condition: the boundaries are set to be 0 to prevent being chosen as the elbow point\\
	for i in range(1, dimension-1):\\
	slope1 = wcss[i] - wcss[i-1]  \# First slope (between wcss[i-1] and wcss[i])\\
	slope2 = wcss[i+1] - wcss[i]  \# Second slope (between wcss[i] and wcss[i+1])\\
	
	\# Check the condition for second slope is smaller (indicating the facing-downwards angle)\\
	if slope2 <= slope1:\\
	tanpsi.append(0)  \# Append 0 to prevent being choses as the elbow point\\
	else:\\
	\# Calculate tan(psi) using the formula for the change in slopes\\
	hasil = (slope1 - slope2) / (1 + (slope2 * slope1))
	tanpsi.append(hasil)\\
	
	\# Append 0 for the last index (boundary condition)\\
	tanpsi.append(0)\\
	
	\# Print the tangents\\
	for i in range(0, len(tanpsi)):\\
	print(f"Tanpsi({i+1}) = {tanpsi[i]}")\\
	
	\# Optimal number of k is the elbow point (the smallest tangent)\\
	optimal\_k = tanpsi.index(min(tanpsi)) + 1\\
	
	print(f"The most optimum number of clusters is {optimal\_k}")\\
\end{algorithm}

\begin{algorithm}
	\caption{Continuation of Algorithm 2}
	kmeans = KMeans(n\_clusters=optimal\_k, init='k-means++', max\_iter=300, n\_init=10, random\_state=0)\\
	y\_kmeans = kmeans.fit\_predict(X)\\
	
	\# Visualize the clusters\\
	plt.scatter(X[y\_kmeans == 0, 0], X[y\_kmeans == 0, 1], s=100, c='red', label='Cluster 1')\\
	plt.scatter(X[y\_kmeans == 1, 0], X[y\_kmeans == 1, 1], s=100, c='blue', label='Cluster 2')\\
	if optimal\_k > 2:\\
	plt.scatter(X[y\_kmeans == 2, 0], X[y\_kmeans == 2, 1], s=100, c='green', label='Cluster 3')\\
	plt.scatter(kmeans.cluster\_centers\_[:, 0], kmeans.cluster\_centers\_[:, 1], s=300, c='yellow', label='Centroids')\\
	plt.title('Clusters of data points')\\
	plt.xlabel('X-axis')\\
	plt.ylabel('Y-axis')\\
	plt.legend()\\
	plt.show()\\
\end{algorithm}

Here is the output of the given implementation on Python.

\begin{figure}[H]
\centering
\includegraphics[scale=0.32]{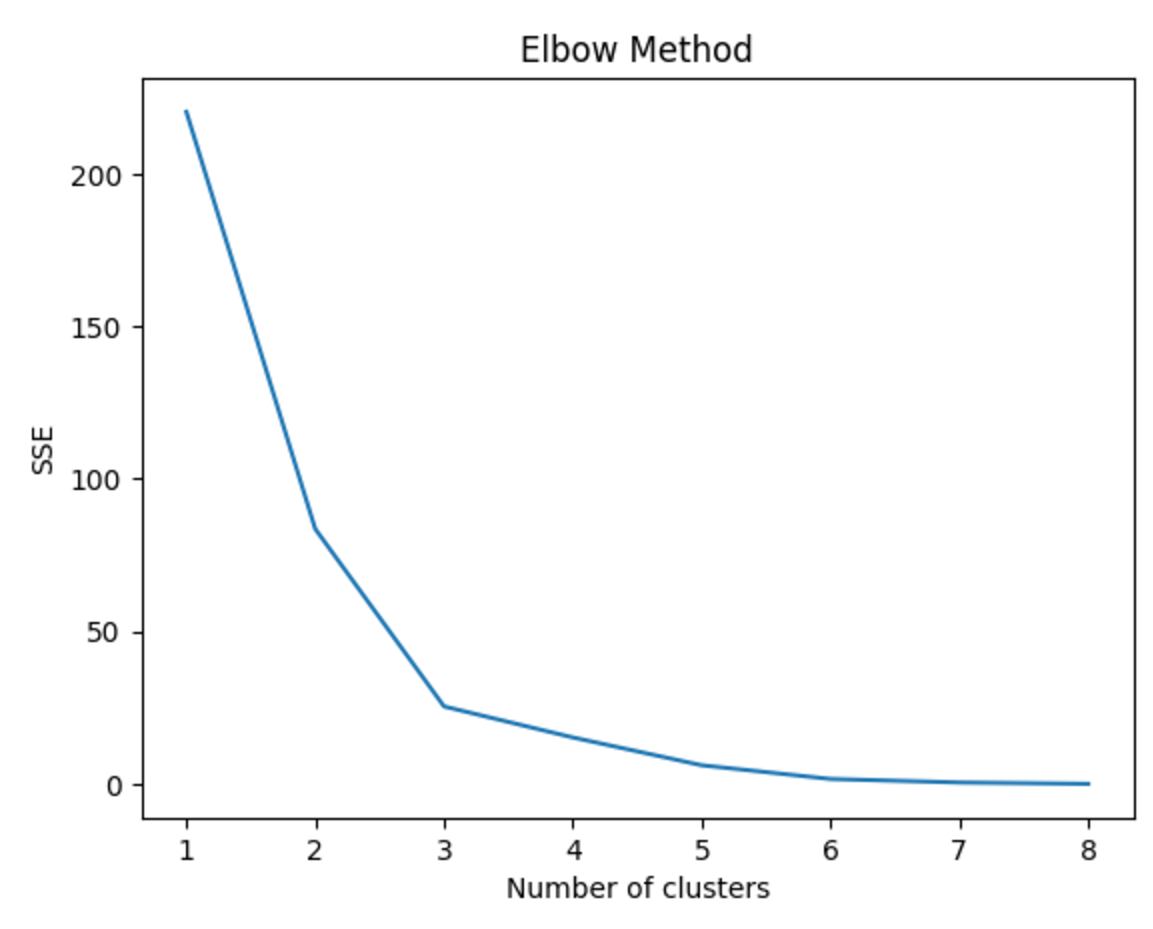}
\caption{SSE plot of Python simulation (distorted)}
\label{fig:elbowpython1}
\end{figure}

\begin{figure}[H]
	\centering
	\includegraphics[scale=0.4]{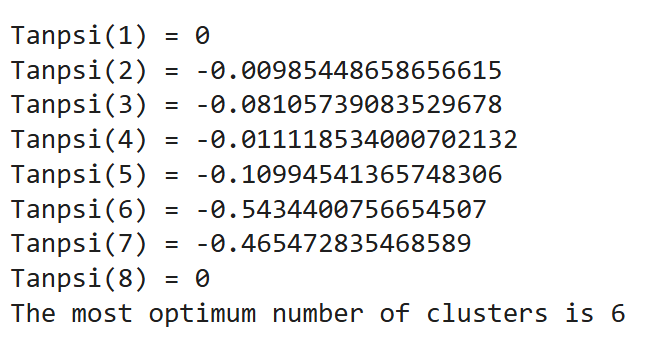}
	\caption{Values of $\tan{\psi_k}$}
	\label{fig:tanpsipython1}
\end{figure}

Figure \ref{fig:elbowpython1} shows that the most optimum number of cluster is $3$. However, as described in Introduction, it could be misleading. Is is answered by Figure \ref{fig:tanpsipython1}.

Figure \ref{fig:tanpsipython1} shows that the smallest tangent is obtained when $k=6$. It shows that the distorted SSE plot in Figure \ref{fig:elbowpython1} is indeed misleading.

\begin{figure}[H]
	\centering
	\includegraphics[scale=0.5]{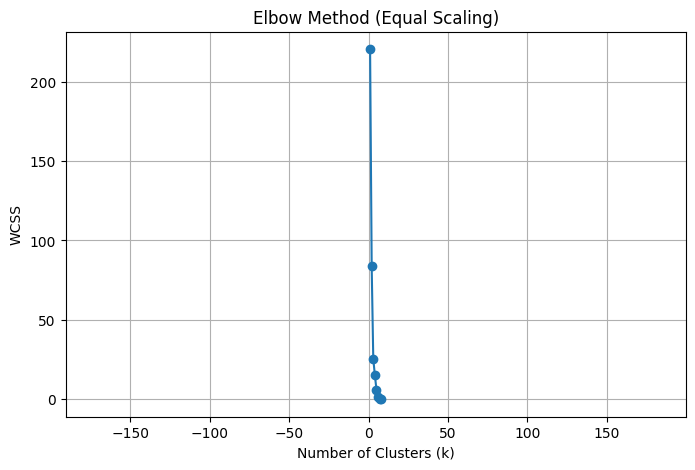}
	\caption{SSE plot of Python simulation (undistorted)}
	\label{fig:elbowpythonfix1}
\end{figure}

\begin{figure}[H]
	\centering
	\includegraphics[scale=0.3]{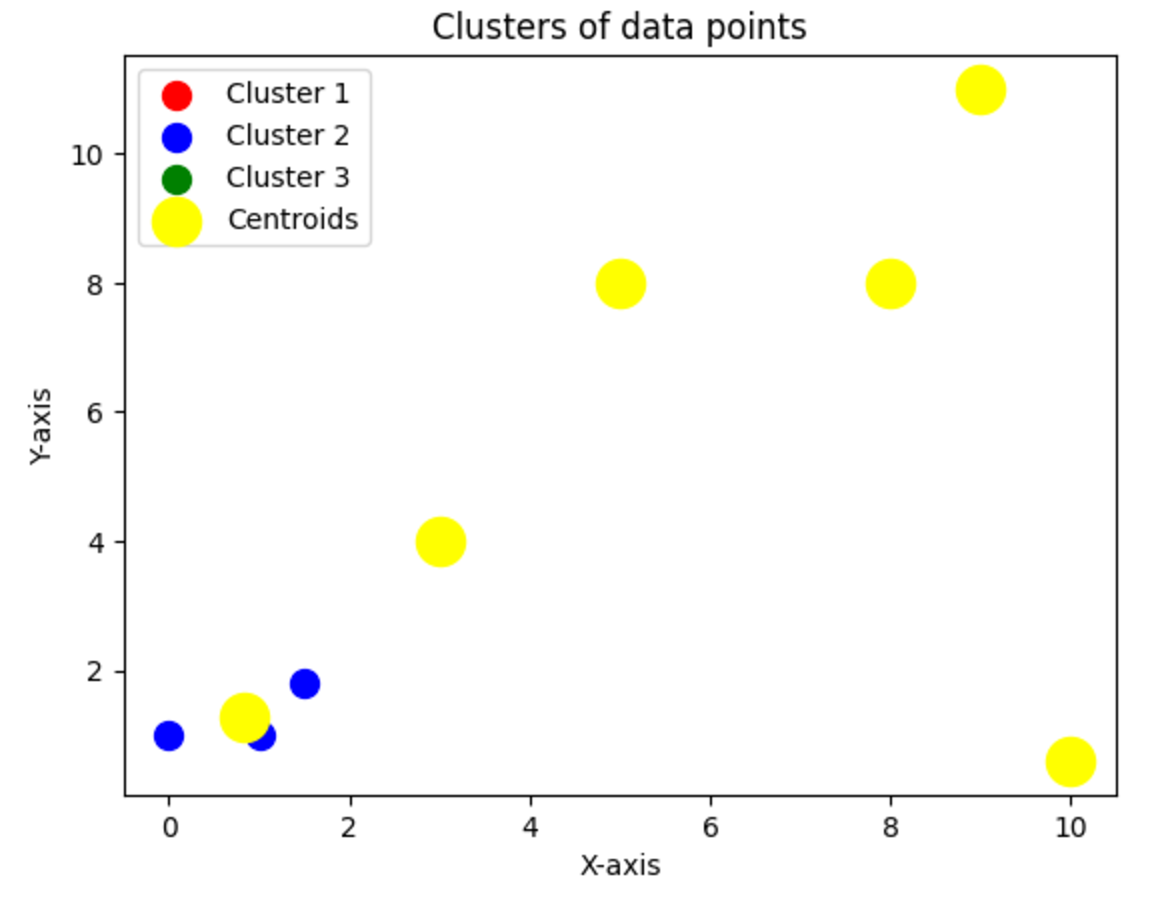}
	\caption{K-means implementation for $k=6$}
	\label{fig:kmeanspython1}
\end{figure}

Figure \ref{fig:elbowpythonfix1} shows the undistorted plot of SSE. Despite the quite unclear plot, we can conclude that $3$ is not the most optimum number of clusters since there is still indeed a significant drop afterwards. The plot starts flattening after $k=6$. The result of the k-means clustering is shown in Figure \ref{fig:kmeanspython1}.

\begin{figure}[H]
	\centering
	\includegraphics[scale=0.63]{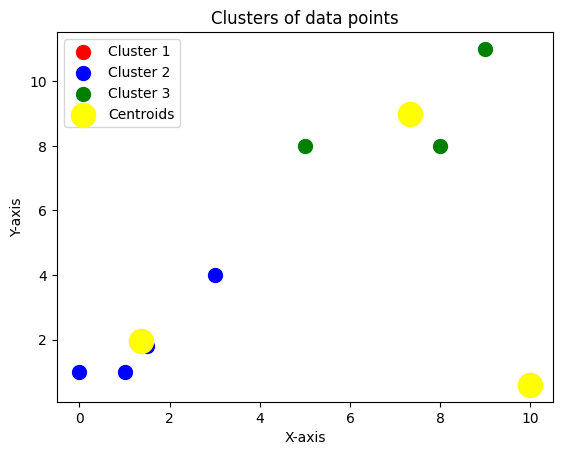}
	\caption{Figure example}
	\label{fig:kmeansbiasedpython1}
\end{figure}

In comparison to Figure \ref{fig:kmeanspython1}, Figure \ref{fig:kmeansbiasedpython1} shows the k-means clustering if there are $3$ clusters, mistakenly chosen by the biased Figure \ref{fig:elbowpython1}. We see that the clustering still leaves a huge distance between data points and their corresponding centroids. Figure \ref{fig:kmeanspython1} shows a better clustering where the distances are minimum.

\vspace{1.5cc}

\section{CONCLUDING REMARKS}

The elbow method, known as a heuristic method, is made exact using the formula of angle between lines. Suppose that $(k,SSE(k))$ are tuple points of number of clusters and corresponding SSE. The elbow point is the point that satisfy
\begin{equation*}
	\min{(\tan{\psi_k} | k=2,3,...,n-1; m_{l_k} > m_{l_{k-1}})}
\end{equation*}
where
\begin{equation*}
	\tan(\psi_k) = \frac{-SSE(k+1)+2SSE(k)-SSE(k-1)}{1+(SSE(k)-SSE(k-1))(SSE(k+1)-SSE(k))}
\end{equation*}
and $m_{l_k} = SSE(k+1) - SSE(k)$ for $k=1,2,...,n$. The algorithm is simple such that it is easily implementable in programming.

\vspace{1.5cc}
\noindent
{\bf Acknowledgement.}We gratefully acknowledge that the research is funded by Institute of Research and Community Service (LPPM), Universitas Jenderal Soedirman, under the contract number 26.713 /UN23.35.5/PT.01/II/2024.

\vspace{2cc}

\bibliographystyle{plain}
\bibliography{elbowref}

\end{document}